\documentclass[
    ,final            
  ]
  {aipproc}

\layoutstyle{8x11single}

\usepackage{color}
\usepackage{epstopdf} 
\usepackage{comment} 

\begin{document}

\title{Photoelectron Emission from Metal Surfaces Induced by VUV-emission of Filament Driven Hydrogen Arc Discharge Plasma}

\classification{29.25.Ni, 52.50.Nr,52.25.Tx}
\keywords      {Negative ion source, arc discharge, photoelectron emission}

\author{J. Laulainen}{
  address={University of Jyväskylä, Department of Physics, Finland}
}
\author{T. Kalvas}{
  address={University of Jyväskylä, Department of Physics, Finland}
}
\author{H. Koivisto}{
  address={University of Jyväskylä, Department of Physics, Finland}
}
\author{J. Komppula}{
  address={University of Jyväskylä, Department of Physics, Finland}
}
\author{O. Tarvainen}{
  address={University of Jyväskylä, Department of Physics, Finland}
}

\begin{abstract}
Photoelectron emission measurements have been performed using a filament-driven multi-cusp arc discharge volume production H$^{-}$ ion source (LIISA). It has been found that photoelectron currents obtained with Al, Cu, Mo, Ta and stainless steel (SAE 304) are on the same order of magnitude. The photoelectron currents depend linearly on the discharge power. It is shown experimentally that photoelectron emission is significant only in the short wavelength range of hydrogen spectrum due to the energy dependence of the quantum efficiency. It is estimated from the measured data that the maximum photoelectron flux from plasma chamber walls is on the order of $1$~A per kW of discharge power.

\end{abstract}

\maketitle



\section{Introduction}
In volume production negative ion sources negative hydrogen ions are produced by electron-molecule and electron-ion collision processes in the plasma discharge~\cite{Bacal2005}. The predominant production channel of negative hydrogen ions is dissociative attachment of a cold electron to a vibrationally excited H$_{2}$ molecule in the ground electronic X$^{1} \Sigma^{+}_{g}$ state:
\begin{equation}
\mathrm{e}_{\mathrm{cold}} + \mathrm{H}_{2}(\mathrm{X}^{1} \Sigma^{+}_{g}; v'') \rightarrow \mathrm{H}_{2}^{-}(^{2}\Sigma^{+}_{u}) \rightarrow \mathrm{H}(1\mathrm{s}) + \mathrm{H}^{-}.
\label{eq:DEA}
\end{equation}
An effective source of vibrationally excited molecules with $v'' > 8$ is radiative decay from singlet states, excited by collisions of ground state molecules with energetic primary electrons~\cite{Nishiura2004}:
\begin{equation}
\mathrm{e}_{\mathrm{hot}} + \mathrm{H}_{2} \rightarrow \mathrm{H}_{2}(\mathrm{B}^{1}\Sigma^{+}_{u} \; \mathrm{or} \; \mathrm{C}^{1}\Pi^{+}_{u}) \rightarrow \mathrm{H}_{2}(\mathrm{X}^{1} \Sigma^{+}_{g}; v'') + h \nu.
\label{eq:radiative_decay}
\end{equation}
Also the plasma chamber walls contribute to negative ion yield by affecting the production of vibrationally excited molecules~\cite{Bacal2005, Bacal2004}. In filament driven arc discharge ion sources the only source of hot electrons capable of ionizing or exciting neutral atoms or molecules from their ground states is thermionic emission from the hot biased filament. Sources of cold electrons required for effective dissociative electron attachment are thermally emitted electrons, whose energy is dissipated in consecutive inelastic collisions, electrons produced by ionization and electrons incident from the plasma chamber walls by secondary electron emission and photoelectron (PE) emission. This study concentrates on the PE emission from the plasma chamber walls by vacuum ultraviolet (VUV) radiation.

\begin{figure}[tb]
  \includegraphics[width=0.95\textwidth]{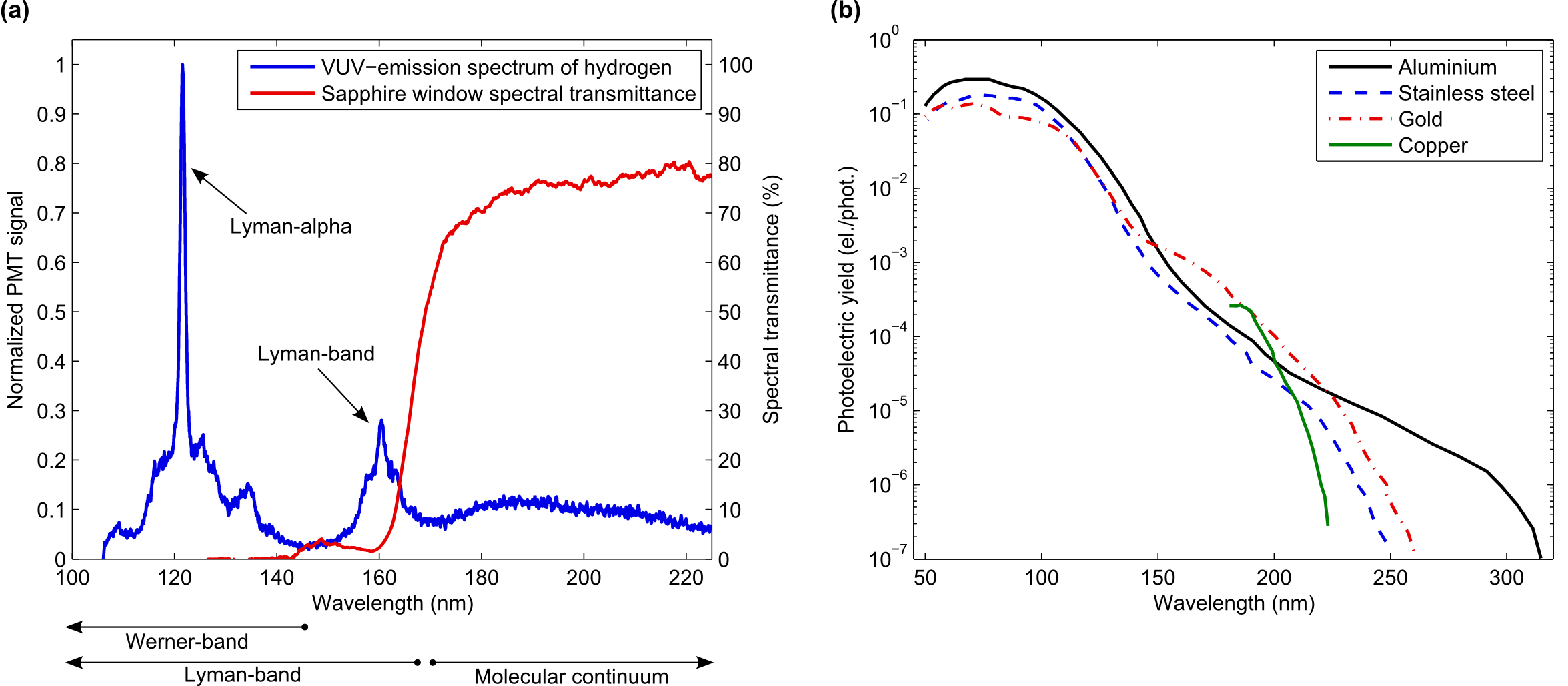}
  \caption{\textbf{(a)} Typical VUV-emission spectrum of the LIISA ion source and the spectral transmittance of the sapphire window. The spectrum is not corrected for spectral transmittance. \textbf{(b)} Photoelectron yield quantum efficiencies for aluminium, stainless steel, gold~\cite{Feuerbacher1972} and copper~\cite{Dowell2006}.}
  \label{fig:spectrum_transmittance_yield}
\end{figure}

Light is emitted by hydrogen plasma as a consequence of electronic transitions from excited states to lower states of neutral atoms or molecules. Typical VUV-emission spectrum from hydrogen plasma is presented in Fig.~\ref{fig:spectrum_transmittance_yield}~(a). The VUV-part of the spectrum is important for PE emission, because for common metals the quantum efficiency increases in the VUV-range with decreasing wavelength of the incident radiation (Fig.~\ref{fig:spectrum_transmittance_yield}~(b)). An intensive source of light in hydrogen plasma is the Lyman-alpha line at $121.6$~nm corresponding to the transition from the first excited state to the ground state (2P $\rightarrow$ 1S) of atomic hydrogen. Lyman-band (B$^{1} \Sigma^{+}_{u} \rightarrow$ X$^{1} \Sigma^{+}_{g}$) at $92$--$184$~nm and Werner-band (C$^{1} \Pi_{u} \rightarrow$ X$^{1} \Sigma^{+}_{g}$) at $84$--$158$~nm correspond to two lowest singlet transitions of hydrogen molecule. The molecular continuum at $165$--$400$~nm corresponds to the lowest triplet transition of molecules (a$^{3} \Sigma^{+}_{g} \rightarrow$ b$^{3} \Sigma^{+}_{u}$). It has been previously measured that in a filament-driven arc discharge $15$--$30$~\% of the discharge power is dissipated via light emission in the VUV-range of $120$--$250$~nm~\cite{Komppula2013}. Taking into account the quantum efficiencies of PE emission this implies that plasma dynamics and their contribution to H$^{-}$ production may be affected by surface processes on plasma chamber walls induced by radiation exceeding the surface work function of the wall material. This paper is dedicated to estimating the maximum PE flux from the plasma chamber walls based on experimental data and discussing the possible effects of the emitted photoelectrons on plasma properties.

\section{Measurement setup}
\begin{figure}[tb]
  \includegraphics[width=0.95\textwidth]{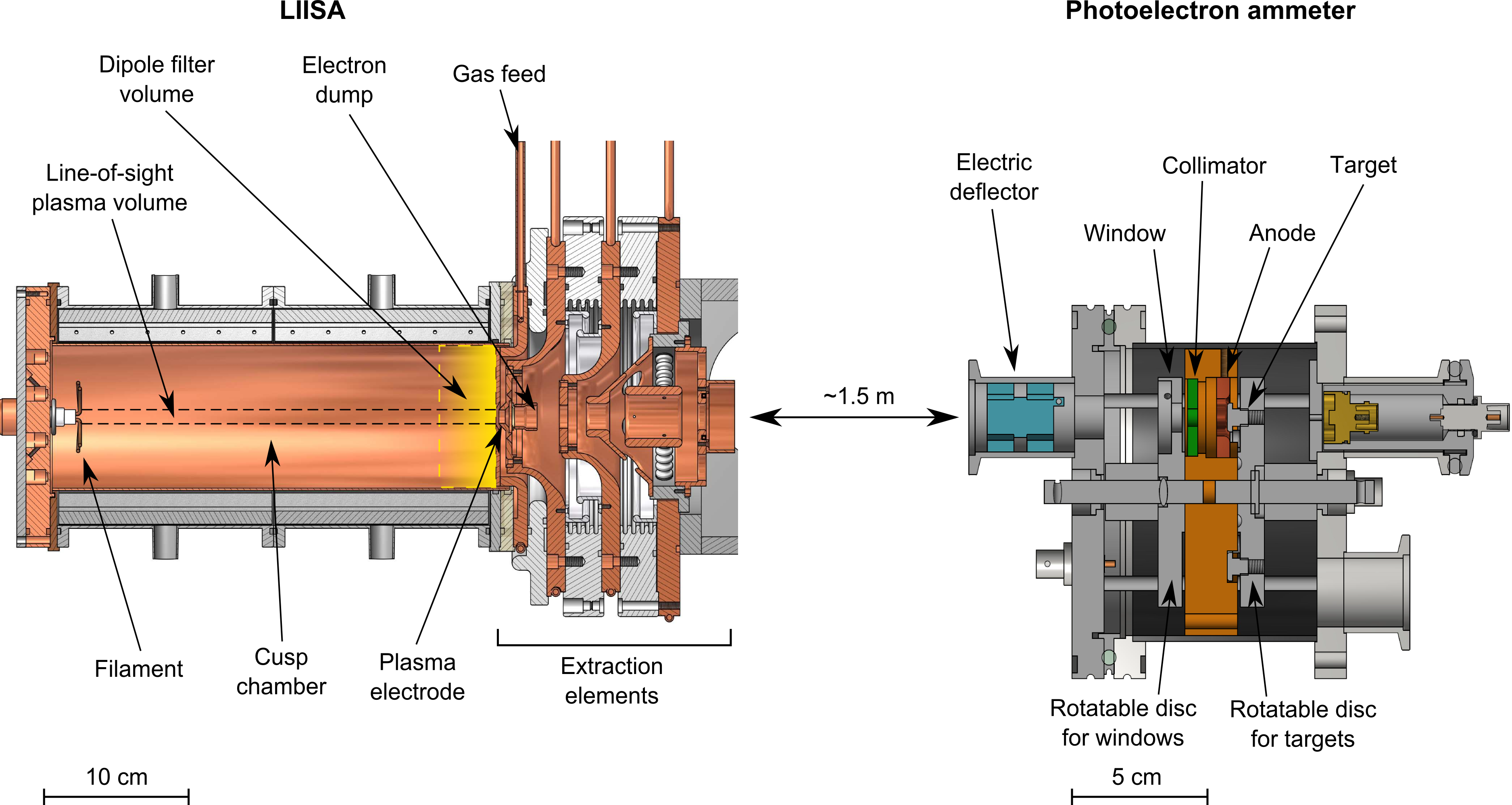}
  \caption{LIISA ion source and the photoelectron current measurement device.}
  \label{fig:setup}
\end{figure}

The experimental setup is presented in Fig.~\ref{fig:setup}. The LIISA (Light Ion Ion Source Apparatus) ion source is a TRIUMF-type DC (tantalum) filament-driven multi-cusp arc discharge volume production ion source. LIISA is designed to provide up to $3$~mA of H$^{-}$ at $5.9$~keV injection energy to the JYFL K130 cyclotron~\cite{Kuo2002}. The typical beam requirement of approximately $1$~mA is reached with $70$~V / $<10$~A discharge voltage / current. The optimum pressure for H$^{-}$ beam current is $3.5 \cdot 10^{-3}$~mbar measured in the plasma chamber. The cylindrical plasma chamber is tantalum-coated copper with $9.8$~cm diameter and $32$~cm length. A cross-sectional view of the ion source is presented in Fig.~\ref{fig:setup}. The PE currents were measured as a function of neutral gas pressure and discharge power. The plasma chamber pressure was varied between $5 \cdot 10^{-4}$~mbar and $1 \cdot 10^{-2}$~mbar and the arc discharge power between $200$~W and $1000$~W by adjusting the arc current and the arc voltage independently. The dependence of the extracted beam current and VUV-emission on these parameters has been reported in Ref.~\cite{Komppula2013}.

The PE currents are measured with a device designed for this work and shown in Fig.~\ref{fig:setup}. The PE ammeter can be equipped with multiple targets and filters mounted to rotatable discs. Different filters can be used to limit the wavelength range of the radiation incident on the target. The PE currents presented in this paper have been measured with an unfiltered view from the plasma to the target and by filtering the incident radiation with a sapphire window. The spectral transmittance of the sapphire window is presented in Fig.~\ref{fig:spectrum_transmittance_yield}~(a) with the VUV-emission spectrum of hydrogen plasma. The sapphire window was used to filter out the short wavelength VUV-radiation in order to demonstrate that the PE emission is significant only in the short wavelength range of hydrogen VUV-emission spectrum. The cathode current is measured from the target with a picoammeter (Keithley 485). The emitted electrons are collected with an anode ring located approximately $3$~mm from the target and biased to $150$~V. The device is protected from plasma particles with an electric deflector, and the light entering the device is collimated with a replaceable collimator. During the measurements the pressure inside the PE ammeter was $\sim 10^{-5}$~mbar. The background signal was measured from a plastic target and with an aluminium plate placed on the rotatable disc housing the filters and, thus, blocking all light. The PE ammeter was placed on the axis of the beamline approximately $1.5$~m from the ion source and was looking into the plasma through the extraction aperture. The line-of-sight plasma volume was limited by the $9.5$~mm diameter puller electrode aperture and the ($6$~mm / $8$~mm diameter) collimator between the target and the plasma. The target materials used in this study are metals which are typically found in negative hydrogen ion sources as chamber materials (Al, Cu, SS)~\cite{Kalvas2013, Kendall1986, Courteille1995}, filament materials (Ta)~\cite{Kuo2002, Kuo1996}, plasma grid materials (Mo)~\cite{Kraus2013} or so-called collar materials (SS, Mo)~\cite{Welton2009, An2010}. The effect of Al, Cu and SS wall materials and Ta covered walls on the volume production of negative hydrogen ions has been studied in~\cite{Bacal2004, Leung1985, Fukumasa1987}.

\section{Experimental results}

\begin{figure}[tb]
  \includegraphics[width=0.5\textwidth]{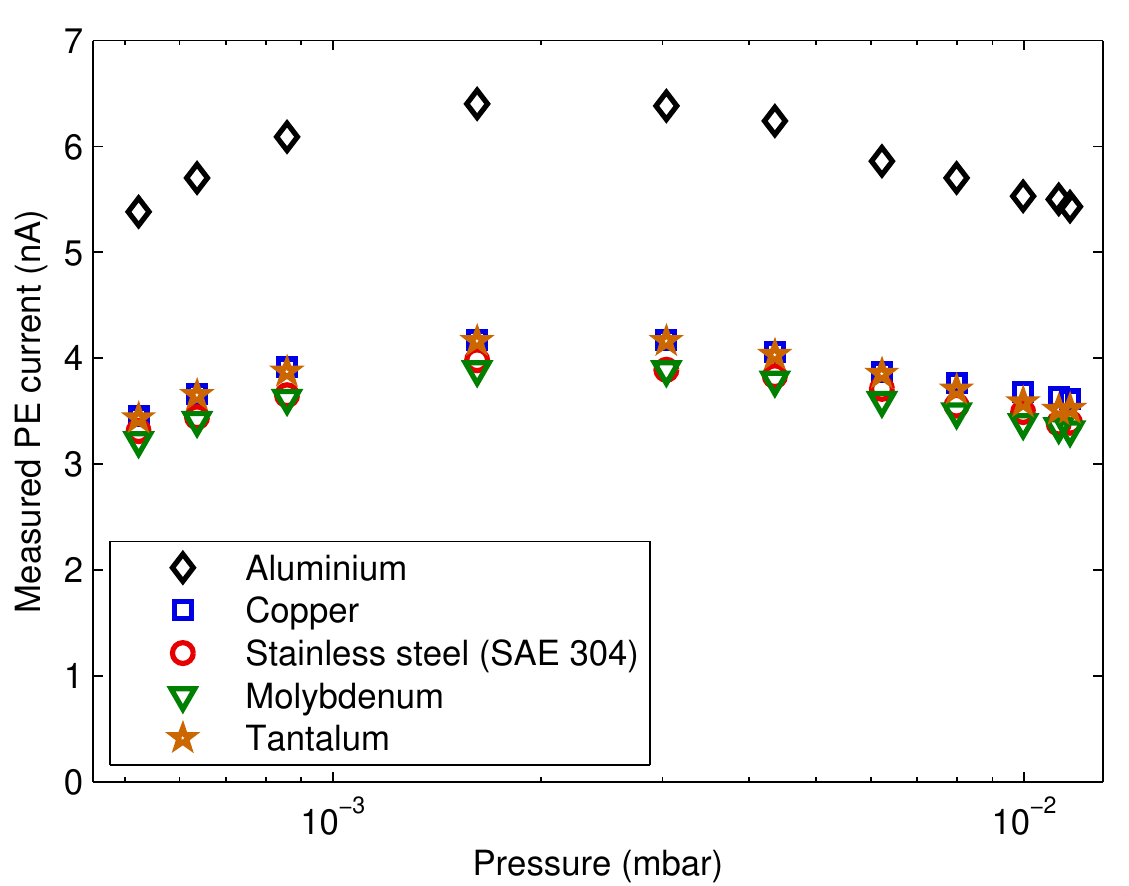}
  \caption{Measured photoelectron currents as a function of neutral hydrogen pressure without filter. The discharge power was $500$~W ($50$~V / $10$~A).}
  \label{fig:pressure}
\end{figure}

The PE currents measured from Al, Cu, stainless steel (SAE 304), Mo, and Ta without filter are presented in Fig.~\ref{fig:pressure}. The PE currents are measured as a function of neutral hydrogen pressure with constant discharge power ($50$~V / $10$~A i.e. $500$~W). The PE currents measured from Cu, SS, Mo, and Ta are approximately equal while the signal from Al is approximately $20$--$50$~\% higher. The dependence of the VUV-light emission on the pressure has been concluded to be weak~\cite{Komppula2013}, and similar behaviour is seen in the PE signals. The change in PE current is less than $20$~\% in the given pressure range. The optimum pressure for PE emission corresponds to the optimum pressure for H$^{-}$ production ($3.5 \cdot 10^{-3}$~mbar).


\begin{figure}[tb]
  \includegraphics[width=1.0\textwidth]{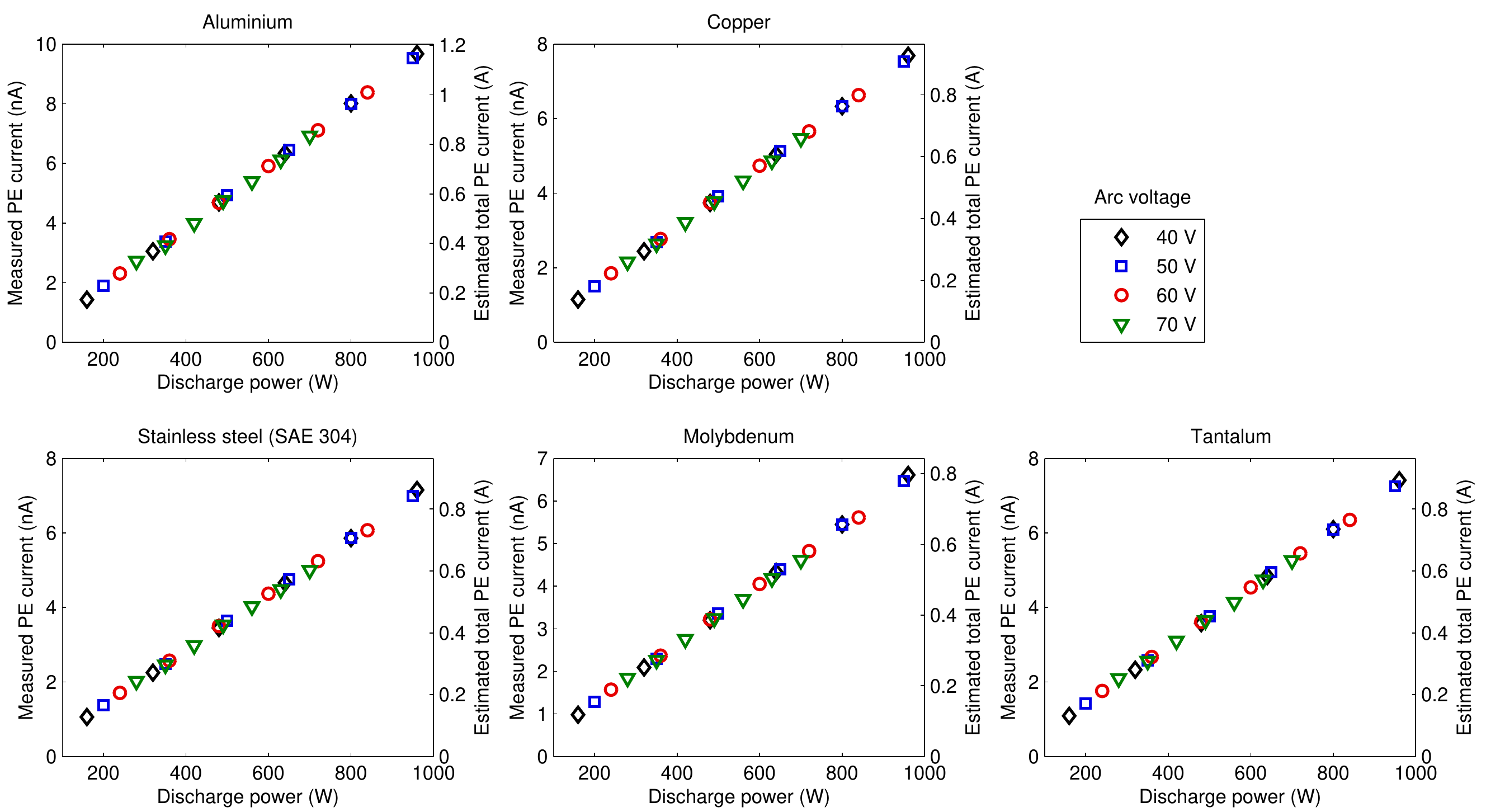}
  \caption{Measured PE currents and estimated total PE currents as a function of discharge current and voltage at $3.8 \cdot 10^{-3}$~mbar pressure without filter.}
  \label{fig:elements}
\end{figure}

The PE currents from different metals measured as a function of discharge power without filter are presented in Fig.~\ref{fig:elements}. The PE currents are measured with various combinations of discharge current and voltage at constant plasma chamber pressure of $3.8 \cdot 10^{-3}$~mbar. The PE currents depend linearly on the discharge power without a preference to neither, the discharge voltage nor current. The Lyman-band and Lyman-alpha light emission has been reported to exhibit similar behaviour. The linear dependence on the discharge power can be explained by the good confinement of hot electrons and their energy dissipation via inelastic collisions as discussed in Ref.~\cite{Komppula2013}. The molecular continuum emission has a notable dependence on the discharge current. However, such trend is not observed in Fig.~\ref{fig:elements}, which implies that the molecular continuum range is insignificant to PE emission from metal surfaces in comparison to shorter wavelength radiation.





\begin{figure}[tb]
  \includegraphics[width=0.5\textwidth]{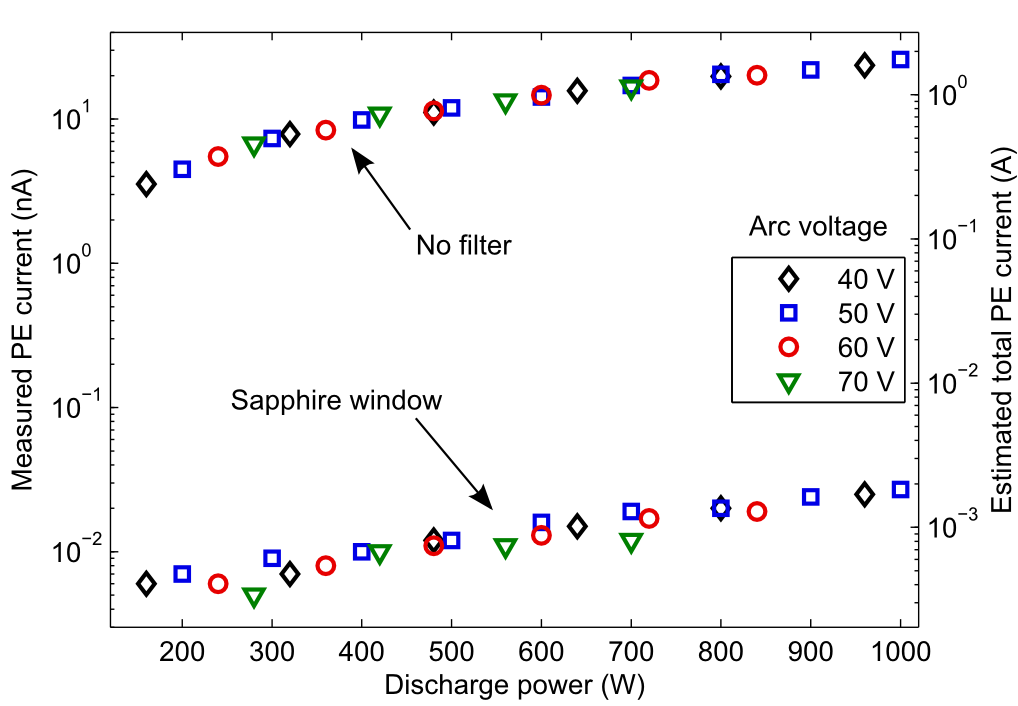}
  \caption{Difference between PE currents from aluminium without filter and with sapphire window at $4.2 \cdot 10^{-3}$~mbar pressure.}
  \label{fig:logplot}
\end{figure}

A comparison between PE currents from Al measured as a function of discharge power without a filter and through a sapphire window at $4.2 \cdot 10^{-3}$~mbar pressure are presented in Fig.~\ref{fig:logplot}. When all wavelengths shorter than $150$~nm are filtered out with the sapphire window, the PE current decreases by three orders of magnitude. Similar behaviour was observed for the other targets as well. The spectral transmittance of the sapphire window (Fig.~\ref{fig:spectrum_transmittance_yield}~(a)) is not $100$~\% at wavelengths longer than $150$~nm, but this has only a small effect on the measured PE current. It is concluded that the decrease in the PE current is due to lower quantum efficiency (Fig.~\ref{fig:spectrum_transmittance_yield}~(b)) at longer wavelengths.

The total PE flux from the plasma chamber walls presented in Figs.~\ref{fig:elements} and~\ref{fig:logplot} is estimated from the measured data by assuming that the light emission profile is homogeneous and isotropic across the plasma chamber profile. Monte Carlo methods were used to calculate the probability for a single photon to reach the target surface. In reality, the spatial distribution of the plasma light emission rate depends on the inhomogeneous plasma density and temperature profiles. Without accurate information about the density and temperature profiles the total PE flux can only be estimated. The assumption used for estimating the total PE flux yields the maximum value. It can be argued that in the line-of-sight volume the plasma density does not change significantly in the axial direction. Thus, the error caused by assuming a constant axial profile is small, because the measured signal corresponds to the average light emission over the axial distance on the line-of-sight. The estimated total PE flux depends on the radial plasma profile because it defines the ratio of observed emission volume to the total emission volume. In the radial direction the plasma density profile is typically almost uniform decaying only near the plasma chamber wall due to the multi-cusp field~\cite{Asano1999, Hwang2006}. The measurement was limited to line-of-sight in the axial direction at the plasma center as illustrated in Fig.~\ref{fig:setup}. Comparing different radial emission profiles to radially homogeneous plasma it can be estimated that the total VUV-emission, and hence the PE current, is at least $50$~\% of the given maximum~\cite{Komppula2013}.

The effect of reflection and scattering of VUV-light from metal surfaces on the measured results can be considered insignificant. In the beam line geometry, scattering is possible only from a small area of the extraction einzel lens or tantalum-coated back plate of the plasma chamber. Also the error due to the picoammeter can be considered insignificant compared to the approximations made for the estimation of the total PE flux.



PE emission is very sensitive to surface contaminants because the penetration depth of VUV-photons and the escape depth of electrons is very short. Furthermore, the surface roughness affects the effective area of the surface, which is directly related to the measured signal. The experimental results are well repeatable after following a standard cleaning procedure of the samples, which includes abrading the surface with sandpaper and wiping with ethanol. Difference between PE currents measured from different Al targets is less than $5$~\%. However, a significant change in the PE current is caused by VUV-induced surface effects. Samples were cleaned in atmospheric pressure, and thus a natural oxide layer was formed on the surface of the metal, which is known to change the work function and the PE quantum efficiency of the surface~\cite{Ramsey1964}. Typical vacuum contaminants, such as residual gases, water vapour and pump oil, are also present on the sample surface. VUV-light is known to destroy organic compounds, in which case free radicals, which can react chemically with the surface, are formed. Consequently, the measurements described in this paper were performed as quickly as possible in order to minimize the effect of VUV-induced surface aging. The change in PE currents due to surface aging was $10$~\% at the most. Conditions and materials used in these experiments are similar to typical ion sources i.e. all phenomena due to surface contaminants are present in ion sources as well. However, in ion sources the surfaces emitting electrons are in direct contact with the plasma which has a cleaning effect that most likely affects the initial conditioning phase of the surfaces and hence the PE emission.

\section{Discussion}
It has been estimated from the measured data that the maximum PE flux from plasma chamber walls is on the order of $1$~A per kW of discharge power. Taking into account the surface area of the LIISA plasma chamber this corresponds to PE emission of $0.9$~mAcm$^{-2}$ per kW on average. Since the PE emission is linearly proportional to the arc discharge power, the numerical results are given normalized to kW of the discharge power. The PE currents obtained with different metals are on the same order of magnitude. The estimated maximum PE current corresponds to almost $10$~\% of the arc current of $14$ A at $70$~V discharge voltage. The arc current consists of thermionic emission of electrons from the filament and the flux of positive ions from the plasma to the filament. The PE flux from the wall to the plasma is limited by the cusp magnetic field, since the cross field diffusion of the emitted electrons in transverse magnetic field is significantly slower than their propagation along the field lines.


The same value for the total PE flux can be obtained independently from the experiments described in this paper by using the measured photon emission rate~\cite{Komppula2013} and known quantum efficiency. The total PE current produced by plasma volume $V$ can be estimated from
\begin{equation}
I \, \mathrm{d}E = \Phi(E) \eta(E) V e \, \mathrm{d}E
\label{eq:pecurrent}
\end{equation}
where $\Phi$ is the volumetric photon emission rate, $\eta$ the quantum efficiency, $V$ the plasma volume, and $e$ the elementary charge. The total VUV-emission power at the Lyman-alpha and Werner-band parts of the spectrum is the most significant and the most emissive part of the spectrum~\cite{Komppula2013}, and the quantum efficiency of common metals is several orders of magnitude higher at short wavelength range of the spectrum (Fig.~\ref{fig:spectrum_transmittance_yield}~(b)). As discussed in the previous section, the PE emission is predominantly caused by radiation at short wavelength range of hydrogen VUV-spectrum. Thus, the total PE emission can be approximated by taking into account only a limited wavelength range. The measured photon emission from LIISA is $\Phi = 5.1 \cdot 10^{16}$ s$^{-1}$cm$^{-3}$ per kW at $122$~nm (FWHM $20$~nm)~\cite{Komppula2013} and the corresponding quantum efficiency of Al at this wavelength is approximately $5 \cdot 10^{-2}$~\cite{Feuerbacher1972}. Substituting the given numbers into Eq.~\eqref{eq:pecurrent} yields approximately $1.0$~A per kW total PE current from the plasma chamber walls, which correlates well with the result reported in this paper.



Based on the VUV-diagnostics it has been deduced that the arc discharge power is well dissipated to the plasma in filament driven arc discharge ion sources with cusp-confinement~\cite{Komppula2013}. It has been concluded that the total path length of hot electrons emitted from the filament is long enough for multiple inelastic collisions with neutrals, which is the main energy dissipation process. This means that only electrons with low energy are escaping the plasma to the plasma chamber surface. For low energy electrons the secondary electron emission yield is small~\cite{Hilleret2000}, and therefore the secondary electron emission can be considered insignificant in filament driven ion sources in comparison to PE emission.




Although the effect of the electrons, emitted from the surfaces, on the total electron density might be considered insignificant~\cite{Drentje2003}, they may have a considerable local effect on plasma properties. Photoelectrons can be beneficial for dissociative electron attachment or they can destroy already existing negative hydrogen ions in e $+$ H$^{-}$ collisions. The role of photoelectrons depends strongly on the plasma parameters and the energy of the emitted electrons, which is determined by the initial kinetic energy of the electrons and by the potential difference in the plasma sheath. Photoelectrons are emitted from metal surfaces with all energies from zero up to the maximum energy~\cite{DuBridge1933}. The maximum energy of the photoelectrons corresponds to the difference between the energy of the absorbed photon and the surface work function. For common metals used in this study the work function is in the order of $4$--$5$~eV~\cite{CRC}. This is for clean surfaces, and the real work function of technical surfaces covered with their natural oxide and contaminants, typically found in ion sources, can be different. The emitted electrons are further accelerated by the plasma sheath potential. For example, the plasma potential in the TRIUMF source has been measured to be about $5$~V~\cite{Hwang2006}. However, extracted negative ions are predominantly created near the (biased) plasma electrode due to short survival length of the negative ions~\cite{Wunderlich2009}. Typically the optimal plasma electrode bias is close to the plasma potential~\cite{Bacal2005}, which minimizes the potential difference across the sheath. If the flux of the electrons emitted from the walls is high enough, the space charge of the emitted electrons (and negative ions) is not fully compensated by incoming positive ions, which can lead into formation of a virtual cathode~\cite{McAdams2012, Amemiya1998}. In surface production H$^{-}$ ion sources the negative ion formation is based on resonant tunneling of electrons from the surface material to hydrogen atoms or protons impinging a metallic surface~\cite{Wunderlich2009}. Surface produced negative ions are retarded by the virtual cathode, and therefore, the virtual cathode could limit the flux of negative ions that can be transported across the sheath and finally extracted.


Surface production efficiency can be significantly enhanced by covering the surface with a thin layer (ideally a sub-monolayer) of cesium, which decreases the work function of the surface. The cesium layer increases the efficiency of other surface processes as well, including PE emission. The effect of PE emission from cesium coated molybdenum surface to the adjacent plasma sheath has been studied by computer simulations~\cite{Wunderlich2011}. It was concluded that the potential in the plasma sheath is not affected by the photoelectrons, but a small potential decrease in the plasma volume was observed. In the simulations~\cite{Wunderlich2011} only molecular continuum radiation was taken into consideration. However, in this paper it is shown that PE emission is predominantly caused by radiation at shorter wavelengths. Thus, it is possible that the PE emission has a greater effect on the sheath structure than previously estimated.

\begin{theacknowledgments}
This work has been supported by the EU 7\textsuperscript{th} framework programme ``Integrating Activities -- Transnational Access'', project number: 262010 (ENSAR) and by the Academy of Finland under the Finnish Centre of Excellence Programme 2012--2017 (Nuclear and Accelerator Based Physics Research at JYFL).
\end{theacknowledgments}



\bibliographystyle{aipproc}   

\bibliography{nibs2014_ref}

\IfFileExists{\jobname.bbl}{}
 {\typeout{}
  \typeout{******************************************}
  \typeout{** Please run "bibtex \jobname" to optain}
  \typeout{** the bibliography and then re-run LaTeX}
  \typeout{** twice to fix the references!}
  \typeout{******************************************}
  \typeout{}
 }

\end{document}